\documentclass[letterpaper,pdftex,rmp,twocolumn,amsmath,amssymb,groupedaddress,floatfix]{revtex4}

\usepackage{graphicx}
\usepackage{natbib}
\usepackage{textcomp}
\usepackage[normalem]{ulem}
\usepackage{makecell}
\usepackage{float}
\usepackage{setspace}
\usepackage[a]{esvect}
\usepackage[font=sf,size=footnotesize,justification=RaggedRight,singlelinecheck=false]{caption}
\usepackage[sf,bf,small]{titlesec}
\titlespacing*{\section}{0pt}{1em}{0em}
\usepackage{placeins}

\usepackage{mathrsfs}
\usepackage{amsmath,amssymb}
\usepackage{bm}

\usepackage{float}

\bibpunct{()}{)}{,}{s}{,}{,}

\usepackage{multirow}
\usepackage{booktabs}
\usepackage{array}

\begin{document}
\makeatletter
\renewcommand\@biblabel[1]{#1.}
\makeatother

\newcommand{\dif}{\mathrm{d}}

\renewcommand{\figurename}{Figure}
\renewcommand{\thefigure}{\arabic{figure}}
\renewcommand{\tablename}{Table}
\renewcommand{\thetable}{\arabic{table}}
\renewcommand{\refname}{\large References}

\addtolength{\textheight}{1cm}
\addtolength{\textwidth}{1cm}
\addtolength{\hoffset}{-0.5cm}

\setlength{\belowcaptionskip}{1ex}
\setlength{\textfloatsep}{2ex}
\setlength{\dbltextfloatsep}{2ex}

\hyphenation{page-rank}

\title{Individual-based approach to epidemic processes on arbitrary dynamic contact networks}


\author{Luis E C Rocha}
\email{luis.rocha@ki.se}
\affiliation{
Department of Mathematics and naXys, Universit\'e de Namur, 8 Rempart de la Vierge, B-5000 Namur, Belgium \\
Department of Public Health Sciences, Karolinska Institutet, 18A Tomtebodav\"agen, S-17177 Stockholm, Sweden}

\author{Naoki Masuda}
\affiliation{Department of Engineering Mathematics, University of Bristol, Woodland Road, Bristol BS8 1UB, United Kingdom}

\begin{abstract}
The dynamics of contact networks and epidemics of infectious diseases often occur on comparable time scales. Ignoring one of these time scales may provide an incomplete understanding of the population dynamics of the infection process. We develop an individual-based approximation for the susceptible-infected-recovered epidemic model applicable to arbitrary dynamic networks. Our framework provides, at the individual-level, the probability flow over time associated with the infection dynamics. This computationally efficient framework discards the correlation between the states of different nodes, yet provides accurate results in approximating direct numerical simulations. It naturally captures the temporal heterogeneities and correlations of contact sequences, fundamental ingredients regulating the timing and size of an epidemic outbreak. Using real-life data, we show that the static network model overestimates the reproduction number but underestimates the infection potential of super-spreading individuals. The high accuracy of our approximation further allows us to detect the index individual of an epidemic outbreak.
\end{abstract}

\maketitle

\noindent 

\section*{Introduction}

Infectious diseases are a major concern of public health because of the potentially high mortality and financial costs to health systems~\cite{Fonkwo2008}. To avoid or reduce the impact of an epidemic outbreak, it is necessary to understand the mechanisms driving the spreading dynamics. The population dynamics of infectious diseases depends on the particular pathogen and on the transmission routes between individuals. Airborne infections, including influenza and tuberculosis, may spread through close contacts between a host and a susceptible individual. Sexual contacts on the other hand create the main route for the spread of infections such as HIV and chlamydia~\cite{KeelingBook2007}. Various forms of daily interactions among people thus form complex contact networks that define the potential infection routes~\cite{Wallinga1999,Danon2011}. These contact networks are characterised by different patterns of connectivity between the individuals and by the timings of the contact events~\cite{Wallinga1999,HolmeSaramaki2012PhysRep}. Previous research has provided substantial understanding of the importance of the structure of contacts (e.g., clustering, number of contacts~\cite{Keeling2005JRSocInterface, May2006, Pastorsatorras2015RevModPhys}) and of temporal correlations (e.g., contact times, concurrency~\cite{Volz2007,Morris1997}) to regulate the spread of infectious diseases. However, how structural and temporal properties compete and synergistically change spreading dynamics remain not sufficiently understood~\cite{Bansal2010, Scholtes2014, Delvenne2015}.

Given the complexity and heterogeneity of contact networks in general, a dominant approach to study epidemics
on dynamic networks is to numerically simulate a stochastic epidemic process. Theoretical approaches however may be useful for further mechanistic understanding of epidemic dynamics and for developing efficient intervention protocols in principled ways. There are several lines of such theoretical approaches \cite{Pastorsatorras2015RevModPhys}. Among them is the individual-based approximation (IBA), also termed discrete-time Markov chains, which is applicable to arbitrary contact networks. The key idea of the IBA is tracking the dynamics of the probability that an individual is in a certain state (e.g., infected state). The IBA has been applied on static networks to study susceptible-infected-susceptible (SIS)~\cite{Wang2003SRDS,Draief2006PhysicaA,Vanmieghem2009IEEETransNetw,Castellano2010PhysRevLett,Gomez2010EPL,Pastorsatorras2015RevModPhys} and susceptible-infected-recovered (SIR)~\cite{Draief2008AnnApplProb,Sharkey2008JMathBiol,Sharkey2011TheorPopulBiol, Youssef2011JTheorBiol,Pastorsatorras2015RevModPhys} epidemic models.
Although the applicability of the IBA is hampered by correlation between the states of different nodes, in particular between adjacent nodes \cite{Sharkey2008JMathBiol,Sharkey2011TheorPopulBiol,Pastorsatorras2015RevModPhys}, the same technique applied to dynamic contact networks may be rather accurate because the dynamics of contact networks may decorrelate the states of different nodes. Individual-based methods have been applied for understanding the physics of the SIS epidemics on dynamic networks, in combination with synthetic adaptive networks \cite{GuoTrajanovski2013PhysRevE} and network data extended with the temporal periodic boundary condition \cite{Valdano2015PhysRevX}.

In the present study, we develop the IBA for the susceptible-infected-recovered model on dynamic contact networks as observed on real settings. Our model describes a broad class of infectious diseases, such as measles, chickenpox, and Ebola, where hosts develop immunity or die after a given infectious period~\cite{KeelingBook2007}. We use our framework to estimate the dynamics of macroscopic epidemiological variables such as the time-dependent prevalence of infections, formulate the effective reproduction number (i.e., the number of secondary infections produced by a single infected individual in a finite population~\cite{Brauer2008}), quantify super-spreaders~\cite{Galvani2005}, and detect the source of infections if past contacts and the epidemiological state of the population are known at a given time~\cite{Fantulin2015}.

\section*{Results}

\noindent\textbf{Dynamic Contact Networks.} A dynamic contact network is defined as a sequence of snapshots. A snapshot is a contact network of $N$ nodes (i.e., individuals) represented by a contact matrix $\bm A(t)=(a_{ij}(t))$ ($1\le i, j\le N$, $1\le t\le t_{\max}$). We set $a_{ij}(t)=1$ if a link (i.e., a contact) $(i, j)$ exists at time $t$ and $a_{ij}(t)=0$ otherwise. We assume that each snapshot is an undirected network such that $\bm A(t)$ is a symmetric matrix. Each snapshot corresponds to time $T_{\rm w}$, representing the temporal resolution of observation of contact networks or the level of coarse graining of contact networks in terms of time. If the total observation time of a contact network is $T$, the number of snapshots is equal to $t_{\max}\equiv T/T_{\rm w}$. The framework of dynamic networks is relevant if $T_{\rm w}$ is smaller than or comparable with the time scale of the epidemics process. Otherwise, the network changes more slowly than the epidemic states, and the static network approximation is sufficient. In the present paper, we study high-resolution contact network data with $T_{\rm w}=20$ sec, which is much smaller than the time scale of real epidemic processes, in which the infectious period typically lasts a few days or more~\cite{NHS2015}.

\vspace{0.5cm}
\noindent\textbf{SIR Model on Dynamic Networks.} We consider the discrete-time SIR model in which a snapshot network corresponds to a single time step. At each time, an individual is either in the susceptible (S), infected (I), or recovered (R) state. Each individual experiences at most one state-transition event within a snapshot. This assumption is valid when $T_{\rm w}$ is sufficiently small relative to the time scale of the SIR dynamics. Upon contact, an infected individual infects a susceptible with (per-contact) probability $\beta$. An infected individual recovers with probability $\mu$ in each snapshot. We assume that the recovery events occur prior to the infection events in each snapshot. This assumption is reasonable if $T_{\rm w}$ is sufficiently small.

\vspace{0.5cm}
\noindent\textbf{Individual-based Approximation.} We denote by $S_i(t)$, $I_i(t)$, and $R_i(t)$ the probability that individual $i$ is in the state S, I, and R at time $t$, respectively; therefore, $S_i(t)+I_i(t)+R_i(t)=1$. The probability $p_{ij}(t)$ that individual $i$ is not infected by individual $j$ at time $t$, under the condition that $i$ is in state S at time $t-1$, is given by
\begin{equation}
p_{ij}(t)=\begin{cases}
\phi_j(t) & (a_{ij}(t) = 1),\\
1 & (a_{ij}(t) = 0),
\end{cases}
\label{eq:p_{ij}(t)}
\end{equation}
for $t\ge 1$, where
\begin{equation}
\phi_j(t) = 1 -  (1-\mu)\beta I_j(t-1).
\label{eq:phi_j(t)}
\end{equation}

If there is no contact between $i$ and $j$ at time $t$, $i$ is not infected by $j$ at this time $t$ such that $p_{ij}(t)=1$. Otherwise, $j$ infects $i$ if and only if $j$ is infected (with probability $I_j(t-1)$), $j$ does not recover at time $t$ (with probability $1-\mu$), and the infection occurs with probability $\beta$ (equation~\eqref{eq:phi_j(t)}). Note that $p_{ij}(t)$ is independent of $i$. 

Equation~\eqref{eq:p_{ij}(t)} is supplied with
\begin{align}
S_i(t) =& S_i(t-1) \prod_{j\in {\cal N}_i(t)} \phi_j(t),
\label{eq:S_i(t)}\\
I_i(t) =& I_i(t-1) + S_i(t-1) \left[ 1 - \prod_{j\in {\cal N}_i(t)} \phi_j(t) \right] - \mu I_i(t-1),
\label{eq:I_i(t)}\\
R_i(t) =& 1 - S_i(t) - I_i(t),
\label{eq:R_i(t)}
\end{align}
where the set of neighbours of the individual $i$ at time $t$ is denoted by ${\cal N}_i(t) \equiv \{j; a_{ij}(t)=1\}$. Equation~\eqref{eq:S_i(t)} and equation~\eqref{eq:I_i(t)} are only approximate because the expression $\prod_{j\in {\cal N}_i(t)} \phi_j(t)$
assumes that $I_j(t)$ for different $j$ values represents independent events. In fact, the states of different individuals are generally correlated. For example, the true probability that two individuals $i$ and $j$ are simultaneously infected at time $t$ may be larger or smaller than $I_i(t) I_j(t)$. It is straightforward to extend the IBA to the case of weighted networks. The IBA is known for the continuous-time SIR model on static networks
\cite{Draief2008AnnApplProb,Sharkey2008JMathBiol,Sharkey2011TheorPopulBiol,Youssef2011JTheorBiol}.
Adapting it to the case of dynamic networks and discretising the time yield a set of equations similar to equations~\eqref{eq:p_{ij}(t)}--\eqref{eq:R_i(t)} (see Supplementary Information).

To calculate the IBA in each time step, we start by calculating $\phi_j(t)$ ($1\le j\le N$), which requires $O(N)$ time. Then, we scan the list of contacts at time $t$, by which we can calculate the most time-consuming part, i.e., $\prod_{j\in {\cal N}_i(t)} \phi_j(t)$. This operation requires $O(N\left<k\right>_{\text{snap}})$ time, where $\left<k\right>_{\text{snap}}$ is the mean number of contacts per individual in a snapshot. Therefore, running the IBA for the entire dynamic network data requires
$O(N\left<k\right>_{\text{snap}} t_{\max})$ time. Running a direct numerical simulation of the SIR dynamics consumes $O(N)$ time for possible recovery events and $O(N\left<k\right>_{\text{snap}})$ time for possible infection events for each time step. Therefore, the total time for a single realisation is of the same order as that for the IBA. The merit of the IBA is thus that it tracks the evolution of the probability, corresponding to infinitely many realisations of direct numerical simulations, with the same order of the computation time.

\vspace{0.5cm}
\noindent\textbf{Accuracy of the Individual-based Approximation.} We calculate $S_i(t)$, $I_i(t)$, and $R_i(t)$ ($1\le i\le N$) in increasing order of time $t$ from $t_1=0$ to $t_2=t_{\max}$. We use each $i$ ($1\le i\le N$) as the index individual such that the initial condition of the IBA is given by $S_j(0) = 1-\delta_{ij}$, $I_j(0)=\delta_{ij}$, and $R_j(0) = 0$, $1\le j\le N$, where $\delta$ is the Kronecker delta. The expected fraction of infected individuals (i.e.\ prevalence) at time $t$ is given by $I(t) = \sum_{i=1}^N I_i(t)/N$. Similarly, the expected fractions of susceptible and recovered individuals at time $t$ are given respectively by $S(t) = \sum_{i=1}^N S_i(t)/N$ and $R(t) = \sum_{i=1}^N R_i(t)/N$.

In the following, we describe the results for the conference network data set \cite{Isella11} (see Materials and Methods). The results for other data sets are qualitatively the same (see Supplementary Information). Figures~\ref{fig:accuracy}(a)--(d) show the evolution of $I(t)$ for two arbitrarily selected index individuals and two sets of epidemiological parameters. In the figure, the estimate by the IBA is compared with other approaches available in the literature, i.e., (i) direct numerical simulation on the dynamic network, abbreviated as S-DN, (ii) numerical simulation on the static version of the network, S-SN, and (iii) the standard well-mixed theory, WMT (see Materials and Methods). In (i) and (ii) we take averages over 200 realisations of the simulation but exclude null outbreaks (i.e. the index individual has infected no other individual in the entire observation interval) to calculate $I(t)$. We exclude null outbreaks because approximations using probabilistic flows, including the IBA, are generally accurate under the condition that minor outbreaks (i.e., realisations of the dynamics causing a small final outbreak size) are eliminated~\cite{Sharkey2008JMathBiol}.

The figure suggests a generally good agreement between the IBA and direct simulations on the dynamic network (S-DN). The IBA captures the effect of variations in the contact patterns within and between multiple days. For example, we observe a second wave of infections after the first wave has decayed to low levels. The fact that multiple waves and sudden changes in $I(t)$ are observed for simulations averaged over 200 realisations indicates that they are phenomena shared by a majority of realisations yielding non-null outbreaks. The waves result from temporal patterns of the networks, i.e., concentration of contacts around certain times of the day and the absence of contacts during night.

\begin{figure}[!htbp]
\begin{center}
\centerline{\includegraphics[scale=1]{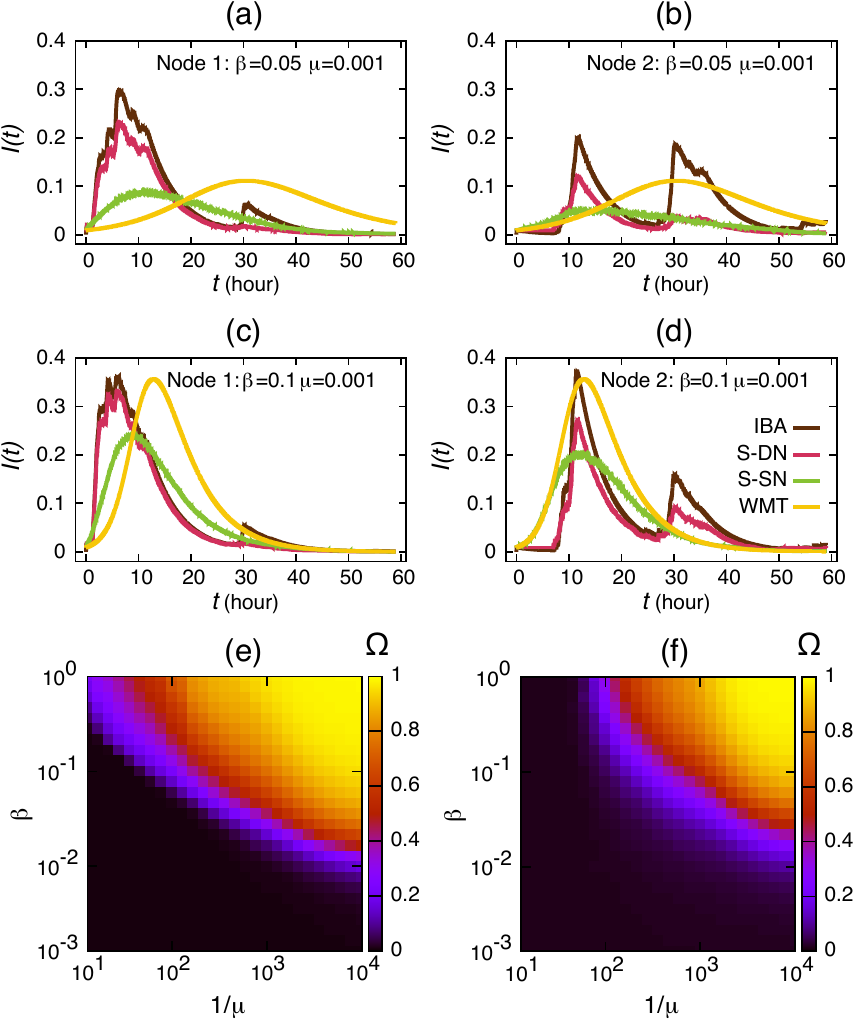}}
\caption{\label{fig:accuracy}\textbf{Accuracy of different approximation methods.} (a)--(d) The fraction of infected individuals as a function of time (i.e.\ the prevalence $I(t)$) for (a, b) $\beta= 0.05$ and $\mu=0.001$, and (c, d) $\beta=0.1$ and $\mu=0.001$, with the infection starting at the seed individual (a, c) 1 or (b, d) 2. (e, f) Final outbreak size $\Omega \equiv R(t_{\max})+ I(t_{\max})$ for (e) the IBA  and (f) the S-DN models. In (e) and (f), we take averages over all individuals as seeds. For the S-DN case, we remove null outbreaks to calculate $\Omega$. The colours represent the values of $\Omega$.}
\end{center}
\end{figure}

The approximate dynamics obtained from the S-SN and WMT are also presented in Fig.~\ref{fig:accuracy}(a)--(d). 
In the S-SN, the temporality of the network is disregarded, and simulations run on the static network generated by aggregating all snapshots with the appropriate normalisation~\cite{Stehle2011}. The WMT disregards both network structure and its temporality. Neither of the two schemes accurately approximates $I(t)$ in the temporal case (i.e., S-DN). In particular, only a single peak is observed with the S-SN and WMT, which contrasts with the results for the S-DN. We also observe that the peak time is shifted with the S-SN and WMT, possibly a result of the dismissal of the burstiness in the timings of contact events as observed in dynamic network models~\cite{Rocha2013PlosComputBiol}.

The static network shows an earlier and higher growth of $I(t)$ ($0 < t < 10$ hours) followed by lower growth than the WMT. This result is possibly because of generally faster spreading in heterogeneous networks due to well-connected individuals (i.e., hubs) in the early stage of the epidemics~\cite{Barthelemy2004PhysRevLett}. We conclude that heterogeneous temporal information substantially affects the time course of epidemic spreading, which is captured by the IBA for temporal networks, but not by methods that disregard temporal information about the contacts.

In a wide region of the $\beta$--$\mu$ parameter space, the final outbreak size $\Omega\equiv R(t_{\max})+I(t_{\max})$ for the S-DN (Fig.~\ref{fig:accuracy}(f)) is accurately estimated by the IBA (Fig.~\ref{fig:accuracy}(e)). The accuracy of the IBA degrades for large $\mu$ because the chance of generating no secondary infections in the simulations increases. The S-SN model tends to overestimate $\Omega$ for larger $\beta$ (see Supplementary Information). The WMT also typically overestimates $\Omega$ for most combinations of $\beta$ and $\mu$ (see Supplementary Information). These results are consistent with those shown in Fig.~\ref{fig:accuracy}(a)--(d) and with previous findings that, at longer times, the SIR epidemic spreads slower on some temporal contacts than on the equivalent static networks~\cite{Karsai2011PhysRevE,Miritello2011PhysRevE,Rocha2013PlosComputBiol, Masuda2013F1000}.

\vspace{0.5cm}
\noindent\textbf{Super-spreading.} Hosts that infect disproportionally more secondary contacts than the average are known as super-spreaders. Super-spreading is observed in a range of infectious diseases such as sexually transmitted infections, SARS, and smallpox~\cite{Galvani2005,Lloydsmith2005Nature} and is not simply determined by the number of contacts that an individual owns but significantly by its position in the contact networks~\cite{Kitsak2010}. Identifying super-spreaders is a fundamental step towards efficient infection interventions because targeting super-spreaders potentially saves resources~\cite{Galvani2005}. We therefore define the individual effective reproduction number~\cite{Lloydsmith2005Nature,CrossLloydsmith2005EcolLett} $R_{\text{eff}}(i, t_1, t_2)$ for dynamic contact networks as
\begin{align}
R_{\text{eff}}(i, t_1, t_2) =& \sum_{t=t_1+1}^{t_2}\sum_{j\in {\cal N}_i(t)}
 S_j(t-1)\left[1-p_{ji}(t)\right]\notag\\
=& \sum_{t=t_1+1}^{t_2} \left[1-\phi_i(t)\right]\sum_{j\in N_i(t)}S_j(t-1),
\label{eq:Reff}
\end{align}
where the epidemic process starts at time $t=t_1$ with the sole infected individual $i$, and $t_2$ is the ending time of the observation. The individual effective reproduction number takes into account the fact that in finite populations, which is the focus of the present study, some individuals may be infected by others before having a chance to be infected by the index individual. We calculate the number of secondary infections caused by $i$ between times $t_1+1 = 1$ and $t_2= t$, and thus abbreviate $R_{\text{eff}}(i, t_1, t_2)$ as $R_{\text{eff}}(i,t)$.

\begin{figure}[!htbp]
\centerline{\includegraphics[scale=1]{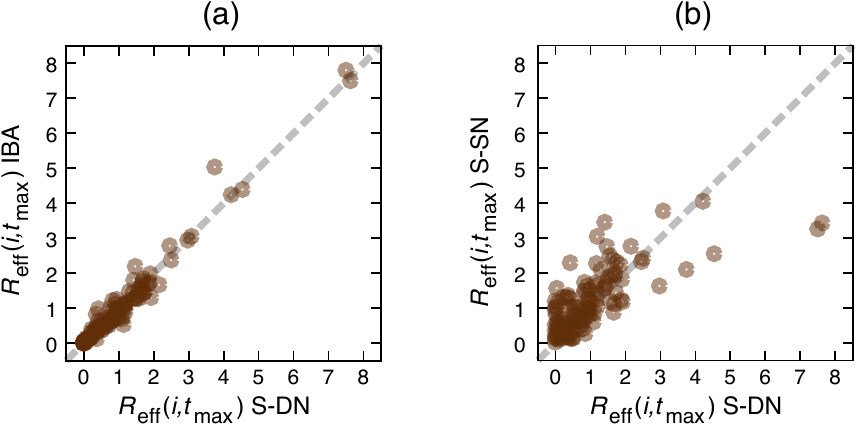}}
\caption{\label{fig02}\textbf{Individual effective reproduction number.} (a) Comparison between the S-DN and IBA. (b) Comparison between the S-DN and S-SN. Each of the $N=113$ circles represents an index individual $i$. Values are estimated at time $t_{\text{max}}$ and the epidemiological parameters are $\beta=0.05$ and $\mu=0.001$.}
\end{figure}

For one pair of $\beta$ and $\mu$ values, $R_{\text{eff}}(i,t_{\max})$ estimated by the IBA is plotted against that calculated by the S-DN in Fig.~\ref{fig02}(a). A circle in the figure represents an index individual $i$. The values for the S-DN represent the number of individuals that $i$ has actually infected, averaged over 200 realisations of the simulation. We first note that the IBA estimates $R_{\text{eff}}(i,t_{\max})$ obtained from the S-DN reasonably well. The static-network approximation, S-SN, is less accurate than the IBA in estimating $R_{\text{eff}}(i,t_{\max})$ (Fig.~\ref{fig02}(b)). Second, the infection potential is highly heterogeneous across individuals. In fact, just a few individuals cause much more secondary infections than the majority of the individuals. The IBA estimates a stronger super-spreading behaviour of some individuals more accurately than the S-SN does (Fig.~\ref{fig02}(b)).

\begin{figure}[!htbp]
\begin{center}
\includegraphics[scale=1]{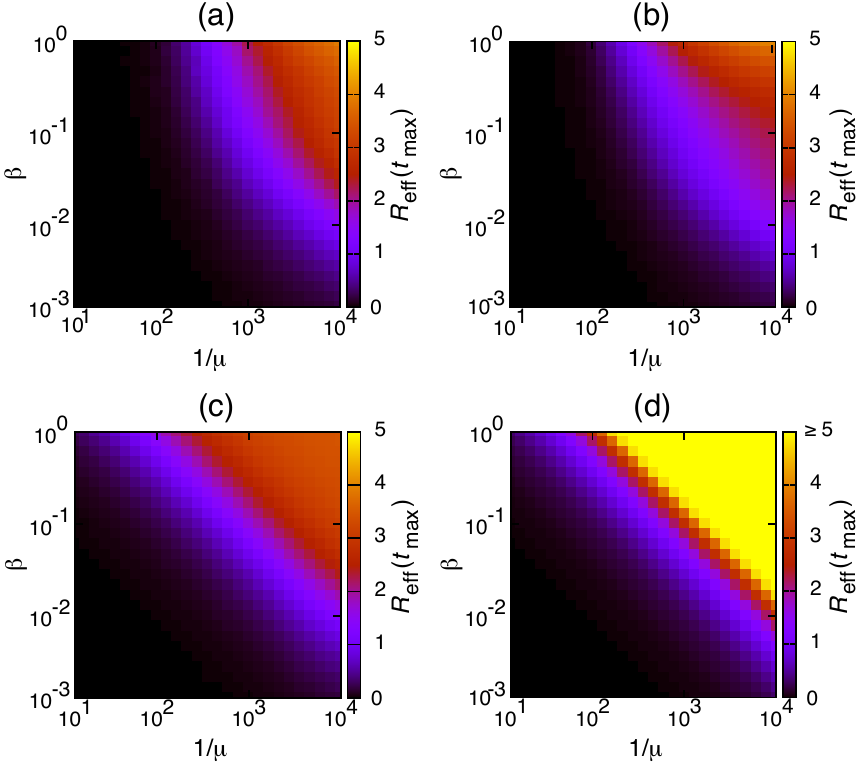}
\caption{\label{fig:effective R0 at t_max}\textbf{Effective reproduction number.} Estimation of the effective reproduction number $R_{\text{eff}}(t_{\max})$ using (a) IBA, (b) S-DN, (c) S-SN, and (d) WMT models, for various epidemiological parameters $\beta$ and $\mu$. The colours represent the value of $R_{\text{eff}}(t_{\max})$. See Materials and Methods for details on the numerical calculations.}
\end{center}
\end{figure}

\vspace{0.5cm}
\noindent\textbf{Effective Reproduction Number.} The basic reproduction number $R_0$, defined as the expected number of secondary infections caused by an index infected individual in a fully susceptible population, is typically used as a threshold to characterise the potential of an epidemic outbreak in the population. The epidemics is likely to occur if and only if $R_0 > 1$~\cite{Heffernan2005, KeelingBook2007}. The reproductive number is a key quantity connected to, for example, the outbreak size and herd-immunity~\cite{Fine2011}. An accurate estimate of $R_{\text{eff}}$ is thus necessary to properly assess the effectiveness of public health interventions. We define the effective reproduction number, $R_{\text{eff}}(t)$, which generalises $R_0$ to be time-dependent and takes into account that the population is not fully susceptible as the time progresses. Using equation~\eqref{eq:Reff}, we define $R_{\text{eff}}(t)$ as the average number of secondary infections caused by a single index individual, i.e., $R_{\text{eff}}(t)= \sum_{i=1}^N R_{\text{eff}}(i,t)/N$. For a range of parameters $\beta$ and $\mu$, $R_{\text{eff}}(t_{\max})$ estimated using the IBA (Fig.~\ref{fig:effective R0 at t_max}(a)) agrees with that obtained from direct numerical simulations, i.e., S-DN (Fig.~\ref{fig:effective R0 at t_max}(b)). The S-SN (Fig.~\ref{fig:effective R0 at t_max}(c)) and WMT (Fig.~\ref{fig:effective R0 at t_max}(d)) generally overestimate $R_{\text{eff}}(t_{\max})$ except when $\beta$ and $\mu$ are both small. Furthermore, exclusively for the S-SN and WMT, contours corresponding to given values of $R_{\text{eff}}(t_{\max})$ look like linear lines in the $\beta$--$\mu$ space.

\begin{figure}[!htbp]
\centerline{\includegraphics[scale=1]{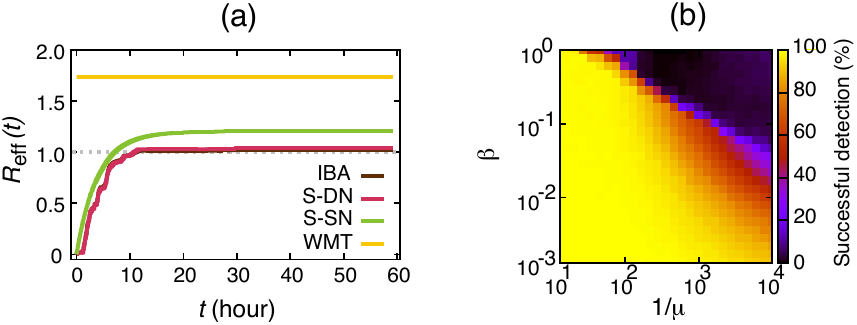}}
\caption{\label{fig04}\textbf{Time dependence of the effective reproduction number and source detection.} (a) Time dependence of the effective reproduction number $R_{\text{eff}}(t)$ calculated using different models with $\beta=0.05$ and $\mu=0.001$. See Materials and Methods for details on the calculations. (b) Percentage of successfully detected infection sources (estimated on the basis of 1000 randomly located sources) for various $\beta$ and $\mu$ values.}
\end{figure}

\vspace{0.5cm}
\noindent\textbf{Importance of Early Times.} Human interaction typically generates strong heterogeneity in temporal contact patterns. The order of the contacts generally regulates the speed and size of an epidemic outbreak since very active individuals may spread the infections quicker than others~\cite{Karsai2011PhysRevE, Rocha2013PlosComputBiol}. The S-SN and WMT models miss this phenomenon because they assume that infection events occur uniformly within the infectious period. Figure~\ref{fig04}(a) shows that $R_{\text{eff}}(t)$ irregularly increases in time if the dynamic contact network is taken into account (i.e., IBA and S-DN). The sudden jumps correspond to the periods in which contacts are dense. We see that $R_{\text{eff}}(t)$ converges within a day, indicating that only early contacts influence the final effective reproduction number, $R_{\text{eff}}(t_{\max})$. The IBA reproduces well the time course of $R_{\text{eff}}(t)$ obtained from the S-DN (brown line in Fig.~\ref{fig04}(a), which overlaps with the red line). Estimation of $R_{\text{eff}}(t)$ on the basis of S-SN (green line in Fig.~\ref{fig04}(a)) also shows saturating behaviour but overestimates $R_{\text{eff}}(t)$ over time. $R_{\text{eff}}(t)$ also builds up too smoothly to explain the behaviour observed for temporal networks. The basic reproduction number $R_0$ for the WMT model is constant in time and shown in Fig.~\ref{fig04}(a) for reference. Our results suggest that as far as $R_{\text{eff}}(t_{\max})$ is concerned, it does not help to sample contact patterns for long times to improve estimation. This reasoning depends on the values of $\beta$ and $\mu$; if $\beta$ or $\mu$ is very small, it takes long time for the epidemics to take off or to get extinguished. Under such conditions, relatively late snapshots may influence $R_{\text{eff}}(t_{\max})$.

\vspace{0.5cm}
\noindent\textbf{Estimation of Source of Infection.} Finding the index case (or patient zero) helps to understand how an infection has been introduced in the population and to trace transmission trees~\cite{Timmreck2002}. The increasing availability of network data has motivated the development of algorithms to detect the source of epidemic spreading on contact networks~\cite{Shah2010SIGMETRICS,Pinto2012PhysRevLett, Brockmann2013Science}. Here we consider the problem of inferring the source of an epidemics when only the information about the current state of each individual and past contact patterns are available~\cite{Fantulin2015}.
Our approach is realistic in the context of schools or hospitals, where contact patterns may be monitored \cite{Salathe2010PNAS, Vanhems2013}. We assume that we can observe the state of all individuals at a given time $t$. We simulate an epidemic outbreak on the contact sequence starting from an infected individual $i_0$. We set a Boolean variable (i.e. 1 if a given state is observed and 0 otherwise) to describe the state of each individual $i$ as susceptible, $N(S, i)$, infected, $N(I, i)$, or recovered, $N(R, i)$ ($1\le i\le N$), such that $N(S, i)+N(I, i)+N(R, i)=1$ (the time variable is suppressed). The aim is to infer the most likely source of infection $\hat{i}_0$ given a configuration of $N(S, i)$, $N(I, i)$, $N(R, i)$ ($1\le i\le N$) at time $t$, and the contact sequence $\bm A(t^{\prime})$ ($t^{\prime}=1, \ldots, t$).

Using the IBA, we calculate $S_i(t)$, $I_i(t)$, and $R_i(t)$ ($1\le i\le N$) for each source individual $i_0$ ($1\le i_0\le N$). For example, $S_i(t)$ is interpreted as the probability that a single realisation yields a configuration at time $t$ such that node $i$ is susceptible (remember that $S_i(t) + I_i(t) + R_i(t) = 1$). The IBA assumes that the states of different individuals are independent of each other. Therefore, the probability that $N(S, i)$, $N(I, i)$, and $N(R, i)$ ($1\le i\le N$) are attained is given by
\begin{equation}
\prod_{i=1}^N S_i(t)^{N(S,i)} I_i(t)^{N(I, i)} R_i(t)^{N(R, i)}.
\label{eq:likelihood}
\end{equation}
The most likely source of infection $\hat{i}_0$ is the $i_0$ value that maximises equation~\eqref{eq:likelihood}.

We set $t=t_{\max}$. Figure~\ref{fig04}(b) shows that the performance of this simple algorithm is relatively good. The fraction of the correct estimation (i.e., $\hat{i}_0 = i_0$) is above $20\%$ for most combinations of $\beta$ and $\mu$. As a comparison, random success would be $1/113 \sim 0.8\%$ for this contact network. The performance of the successful detection degrades for large infection probabilities (e.g., $\beta>0.1$) and long infection periods (e.g., $1/\mu > 1000 \sim 5.55$ hours), where almost the entire population gets infected (see Fig.~\ref{fig:accuracy}(e)). In contrast, the source detection is successful in more than $80\%$ of the cases if the size of the epidemics is not too large (see Fig.~\ref{fig:accuracy}(e)).

\section*{\large Discussion}

The occurrence of epidemic outbreaks typically depends on the pathogen and contact patterns between hosts and susceptible individuals. Previous studies emphasised the importance of the structure of the contact networks on the infection dynamics. The increasing availability of high-resolution longitudinal data has shown, however, that contact patterns are also dynamic across various time scales. A single framework, able to capture all these structural and temporal patterns, is thus necessary to fully understand the population dynamics of infectious diseases.

In this paper, we presented the individual-based approximation (IBA) to model SIR epidemics on arbitrary dynamic contact networks. The IBA neglects the correlations between the states of different individuals, most significantly between neighbouring network nodes. For example, it misses the fact that an individual is more likely to be infected if its neighbour is infected and vice versa. The correlation would develop for networks with a small mean number of contacts and high clustering (i.e., three nodes forming a connected triangle). In such a case, the pair approximation, which explicitly tracks the evolution of the probabilities of pairwise states (the states of adjacent pairs of individuals) at a mean field level \cite{Keeling2005JRSocInterface, Pastorsatorras2015RevModPhys}, or an extension of the IBA to account for pairwise correlation~\cite{Sharkey2008JMathBiol,Sharkey2011TheorPopulBiol} is more accurate. The concept of pairwise correlation is less straightforward if nodes and links appear and disappear in time. The temporality of the network may effectively decorrelate the state of the individuals, which may be why the IBA is relatively accurate at approximating the results obtained from direct numerical simulations.

The IBA showed better performance than well-mixed and static network models to estimate the final outbreak size and the reproduction number on real-life dynamic contact networks. The temporal framework revealed that super-spreaders were more likely than expected from the non-temporal models. These results indicate that the timings of contacts cannot be discarded if one wants to estimate epidemic outbreaks. Our results further suggest that only relatively short intervals of contact network data are necessary to estimate the reproduction number. More importantly, we showed that if longitudinal network data were available, the source of an epidemic could be efficiently detected, assuming a SIR dynamics, even if only the current state of the epidemics was known. These illustrative applications of the IBA may be used for surveillance in closed environments, such as schools~\cite{Salathe2010PNAS} and hospitals~\cite{Vanhems2013}, where protocols to collect human interaction data are already available and outbreaks of infectious diseases may have major consequences. Furthermore, since the IBA framework provides a principled way to calculate probability flows with a single sweep of a given contact network, it allows a computationally efficient estimation of the most-likely transmission trees on large dynamic networks. This information can be exploited to design efficient strategies for infection control such as immunisation~\cite{Bootsma2006} and travel restrictions~\cite{Hollingsworth2006NatMed,BajardiPoletto2011PlosOne}, or to identify sentinels for early detection of epidemics~\cite{Christakis2010, Bajardi2012JRSocInterface}.

To calculate the prevalence for the numerical simulations on dynamic networks (i.e., S-DN), we have excluded the realisations yielding no secondary infections~\cite{Sharkey2008JMathBiol}. This means that the actual average prevalence would be lower than the results shown in our figures, possibly worsen the agreement between the IBA and S-DN. To extend the IBA to account for the absorption probability (epidemic spreading terminates at time $t$) may be a useful improvement. It is not difficult to calculate the probability that no secondary infection happens. 

The present study is limited to SIR epidemics, which is not necessarily the optimal model for various infectious diseases, and to relatively small contact networks, missing, for example, seasonal variations in human interactions. Future studies should investigate the generalisability of our framework to other epidemiological scenarios, particularly taking into account infections with time scales larger than the ones studied here, such as measles and mumps.

\begin{footnotesize}
\section*{\large Methods}
\noindent

We provide a description of the data set used for the analysis in the main text, the SIR model for the well-mixed theory, and the protocol for the numerical simulations for static and dynamic networks. In the Appendix, we derive the discretisation of the continuous-time IBA, present further analyses to support the accuracy of the IBA, including a comparison of the final outbreak size, and analyse other data sets to support our main conclusions. \\

\noindent\textbf{A. Network Data.} \label{SI_sec1}We use a data set of dynamic contact networks representing face-to-face human interactions between delegates in a conference~\cite{Isella11}. Each of the $N=113$ individuals wore a wireless device such that a close face-to-face event was recorded every $T_{\rm w}=20$ sec. The conference lasted for $T\sim 59$ hours, which gave $t_{\max}=10,618$. The time of the first contact defines $t=0$. The data set has $C=20,818$ contacts between $E=2,196$ unique pairs of individuals. The other data sets and the corresponding results are described in the Supplementary Information.\\

\noindent\textbf{B. Well-mixed Theory.} In the WMT case, we solve the following set of equations for the SIR epidemics:
\begin{align}
S(t) =& S(t-1) - \beta' S(t-1) I(t-1), \\
I(t) =& I(t-1) + \beta' S(t-1) I(t-1) - \mu R(t-1),
\label{eq:wm} \\
R(t) =& 1 - S(t) - I(t),
\end{align}
with the infection rate given by $\beta' = \theta \beta$, where $\theta=2C/(t_{\max}N)$ is the contact rate and $\beta$ is the per-contact infection probability. The initial conditions are given by $S(0)=1-I(0)$, $I(0)=1/N$ and $R(0)=0$. We define the effective reproduction number as $R_{\text{eff}}=\beta'/\mu S(0)$.\\

\noindent\textbf{C. Simulations on Static and Dynamic Networks.} For the S-SN, we simulate the epidemic process directly on the static network. First, we build an unweighted network in which a link exists if the contact occurs on the link at least once in $1\le t\le t_{\max}$. We then assign the infection rate $\beta'' = \phi_{ij} \beta$ to each link, where $\phi_{ij}= \sum_{t=1}^{t_{\max}} a_{ij}(t)/t_{\max}$, such that the weight of link $(ij)$ for the dynamic network averaged over time is equal to that for the static network~\cite{Stehle2011}. At each time step, each infected individual may recover with probability $\mu$ and then we check for potential infection events. The algorithm for the dynamic network (i.e., S-DN) is similar. At each time step, an infected individual may recover with probability $\mu$ and then may infect its active contacts with probability $\beta$. In both models, we then update the state of all nodes, measure the macroscopic epidemiological variables, and move to the next time step. For each infection seed, we take averages over 200 realisations.

To calculate the effective reproduction number using direct numerical simulations~\cite{Rocha2013PlosComputBiol, HolmeMasuda2015PlosOne}, we start the infection at a given node (i.e.\ the index individual) and set the rest of the population to susceptible state. We then let the infection evolves, as described above, and count the number of secondary infections caused by this index individual between times $t_1+1$ and $t_2$. For each index individual, we take the average over 200 realisations as the numerical estimation of $R_{\text{eff}}(i, t_1, t_2)$.

\section*{\large Acknowledgements} LECR is a Charg\'e de recherche of the Fonds de la Recherche Scientifique - FNRS. NM acknowledges the support provided through CREST, JST.

\section*{\large Author Contributions} LECR and NM conceived the project, performed the analysis and wrote the manuscript.

\end{footnotesize}

\section*{Appendix}
\section*{Discretisation of the continuous-time IBA}

The IBA to the SIR dynamics on static networks in continuous time has been known~\cite{Sharkey2008JMathBiol, Sharkey2011TheorPopulBiol,Youssef2011JTheorBiol}. By translating it to dynamic contact networks, we obtain the following approximate deterministic dynamics in continuous time:
\begin{align}
\frac{{\rm d}S_i(t)}{{\rm d}t} =& - S_i(t) \beta^{\prime} \sum_{j\in {\cal N}_i(t)}I_j(t),
\label{eq:dS_i/dt individual-based meanfield}\\
\frac{{\rm d}I_i(t)}{{\rm d}t} =& S_i(t) \beta^{\prime} \sum_{j\in {\cal N}_i(t)}I_j(t) - \mu^{\prime} I_i(t),
\label{eq:dI_i/dt individual-based meanfield}\\
\frac{{\rm d}R_i(t)}{{\rm d}t} =& \mu^{\prime} I_i(t),
\label{eq:dR_i/dt individual-based meanfield}
\end{align}
where $\beta^{\prime}$ and $\mu^{\prime}$ are the infection and recovery rates, respectively. We distinguish them from the infection probability $\beta$ and the recovery probability $\mu$ in the main text because we assume discrete time in the IBA, whereas Equations~\eqref{eq:dS_i/dt individual-based meanfield},
\eqref{eq:dI_i/dt individual-based meanfield}, and \eqref{eq:dR_i/dt individual-based meanfield} assume continuous time.

We discretise equations~\eqref{eq:dS_i/dt individual-based meanfield}, \eqref{eq:dI_i/dt individual-based meanfield}, and \eqref{eq:dR_i/dt individual-based meanfield} with a time step $\Delta t$ to obtain
\begin{align}
S_i(t) =& S_i(t-\Delta t) - S_i(t-\Delta t) (\beta^{\prime} \Delta t) \sum_{j\in {\cal N}_i(t-\Delta t)}I_j(t-\Delta t),
\label{eq:dS_i/dt individual-based meanfield Deltat = 20}\\
I_i(t) =& I_i(t-\Delta t) + S_i(t-\Delta t) (\beta^{\prime} \Delta t) \sum_{j\in {\cal N}_i(t-\Delta t)}I_j(t-\Delta t)\notag\\
&- (\mu^{\prime} \Delta t) I_i(t-\Delta t),
\label{eq:dI_i/dt individual-based meanfield Deltat = 20}\\
R_i(t) =& R_i(t-\Delta t) + (\mu^{\prime} \Delta t) I_i(t-\Delta t).
\label{eq:dR_i/dt individual-based meanfield Deltat = 20}
\end{align}
The time discretisation is justified when the probabilities of state-transition events within $\Delta t$ are sufficiently small.
In the present case, this is equivalent to saying $\beta^{\prime}\Delta t, \mu^{\prime}\Delta t \ll 1$. By assuming that $\Delta t$ is the duration of the single snapshot of temporal networks, we obtain $\beta = \beta^{\prime}\Delta t$ and $\mu = \mu^{\prime}\Delta t$. Because our discrete-time approach is justified only when $\beta, \mu\ll 1$, the time discretisation of the continuous-time SIR model is consistent with the assumption justifying our discrete-time approach.

\begin{figure}[h]
\centerline{\includegraphics[scale=1.3]{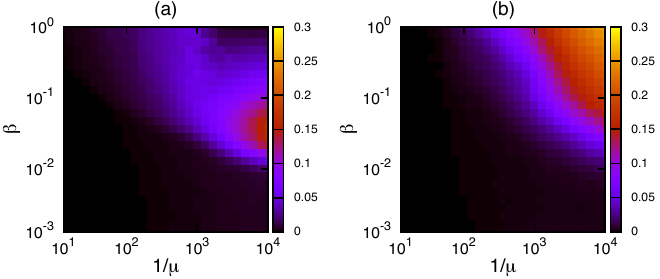}}
\caption{\label{figS01}The root-mean-square RMS (equation~\eqref{eq:01}) between (a) the IBA and S-DN and (b) the S-SN and S-DN for various combinations of infection probability $\beta$ and recovery probability $\mu$ for the SPC data set. The colours represent the value of the RMS.}
\end{figure}

To make an intuitive understanding and comparison with the IBA easier, we change from the continuous-time to discrete-time notation and replace $t-\Delta t$ by $t-1$:
\begin{align}
S_i(t) =& S_i(t-1) - S_i(t-1) \beta \sum_{j\in {\cal N}_i(t-1)}I_j(t-1),
\label{eq:dS_i/dt individual-based meanfield discrete time}\\
I_i(t) =& I_i(t-1) + S_i(t-1) \beta \sum_{j\in {\cal N}_i(t-1)}I_j(t-1)\notag\\
& - \mu I_i(t-1),
\label{eq:dI_i/dt individual-based meanfield discrete time}\\
R_i(t) =& R_i(t-1) + \mu I_i(t-1).
\label{eq:dR_i/dt individual-based meanfield discrete time}
\end{align}

\begin{figure}[h]
\centering
\includegraphics[scale=0.95]{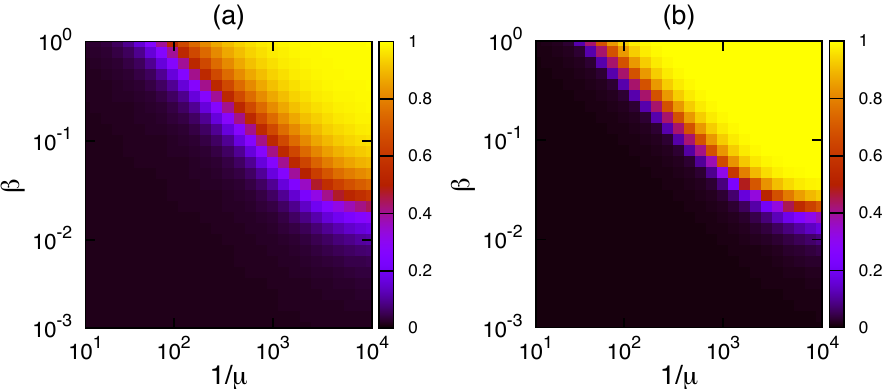}
\caption{\label{figS02}Final outbreak size $\Omega \equiv R(t_{\max})+ I(t_{\max})$ for (a) S-SN and (b) WMT models calibrated with the SPC data set. The colours represent the value of $\Omega$.}
\end{figure}

If we expand equations~(3) and (4) in the main text in terms of small parameters $\beta$ and $\mu$, and only retain the first-order terms, we obtain the variants of equations~\eqref{eq:dS_i/dt individual-based meanfield discrete time} and \eqref{eq:dI_i/dt individual-based meanfield discrete time}, where ${\cal N}_i(t-1)$ in equations~\eqref{eq:dS_i/dt individual-based meanfield discrete time} and \eqref{eq:dI_i/dt individual-based meanfield discrete time} is replaced by
${\cal N}_i(t)$. This minor difference arose because, in equations~\eqref{eq:dS_i/dt individual-based meanfield Deltat = 20}
and \eqref{eq:dI_i/dt individual-based meanfield Deltat = 20}, we used the snapshot network at $t-\Delta t$ to evolve the dynamics, whereas the snapshot network at $t$ is used to evolve the dynamics in equations~(3) and (4) in the main text.

\section*{Accuracy of the IBA}

To estimate the difference in the prevalence obtained from different models, we calculate the root-mean-square (RMS) distance given by
\begin{equation}
\label{eq:01}
RMS = \sqrt{ \sum_{i=1}^{N} \sum_{t=1}^{t_{\max}} \dfrac{[\Omega_{M1}(i,t)-\Omega_{M2}(i,t)]^2}{Nt_{\max}} }
\end{equation}

where M1 and M2 are either the IBA, S-DN, or S-SN, and $\Omega(i,t)$ is the average prevalence at time $t$ when the epidemic spreading starts from a population in which individual $i$ is infected and the other $N-1$ individuals are susceptible. Figure~S\ref{figS01}(a) shows that the RMS between the IBA and S-DN is small for nearly all combinations of $\beta$ and $\mu$. Figure~S\ref{figS01}(b) indicates that the RMS between the S-DN and S-SN is larger in a range of values of $\beta$ and $\mu$. These findings indicate that the results shown in Figure 1(a)-(d) in the main text generally hold true for other combinations of $\beta$ and $\mu$.

\section*{Final outbreak size for S-SN and WMT}

We show the final outbreak size ($\Omega_{t_{\max}}$) for the simulation on static network (S-SN) and the well-mixed theory (WMT) models in Figs.~S\ref{figS02}(a) and S\ref{figS02}(b), respectively. These results should be compared with Figures~1(e) and~1(f) in the main text, respectively, for the individual-based approximation (IBA) and simulation on dynamic networks (S-DN). In contrast to the IBA, both S-SN and WMT do not accurately approximate the results obtained from the S-DN for nearly all values of $\beta$ and $\mu$ considered. The S-SN and WMT agree with the S-DN only for the combination of large $\beta$ combined with and small $\mu$ values.

\begin{table}
\caption{\label{tabS01}Summary of the contact network data sets.}
\begin{tabular}{ccccccc}
\hline
          & $N$\tablenote{Number of individuals ($N$), number of contacts ($C$), number of unique pairs of contacts ($E$), recording time ($T$), number of snapshots ($T_{\text{w}}$), and the average contact rate ($\theta = 2C(TN)^{-1}$).}   & $C$      & E           & $T$ (hour)  & $T_{\text{w}}$ (sec) & $\theta$ \\
\hline
SPC   & 113   & 20,818   & 2,196  & 58.99         & 20                & 0.0347 \\
SPM   &  72   & 6,980     & 691      & 7.29          & 20               & 0.1479 \\
SPH   &   75   & 32,424  & 1,139   & 96.57        & 20                & 0.0497 \\
\hline
\end{tabular}
\end{table}

\begin{figure}[h]
\centering
\includegraphics[scale=0.65]{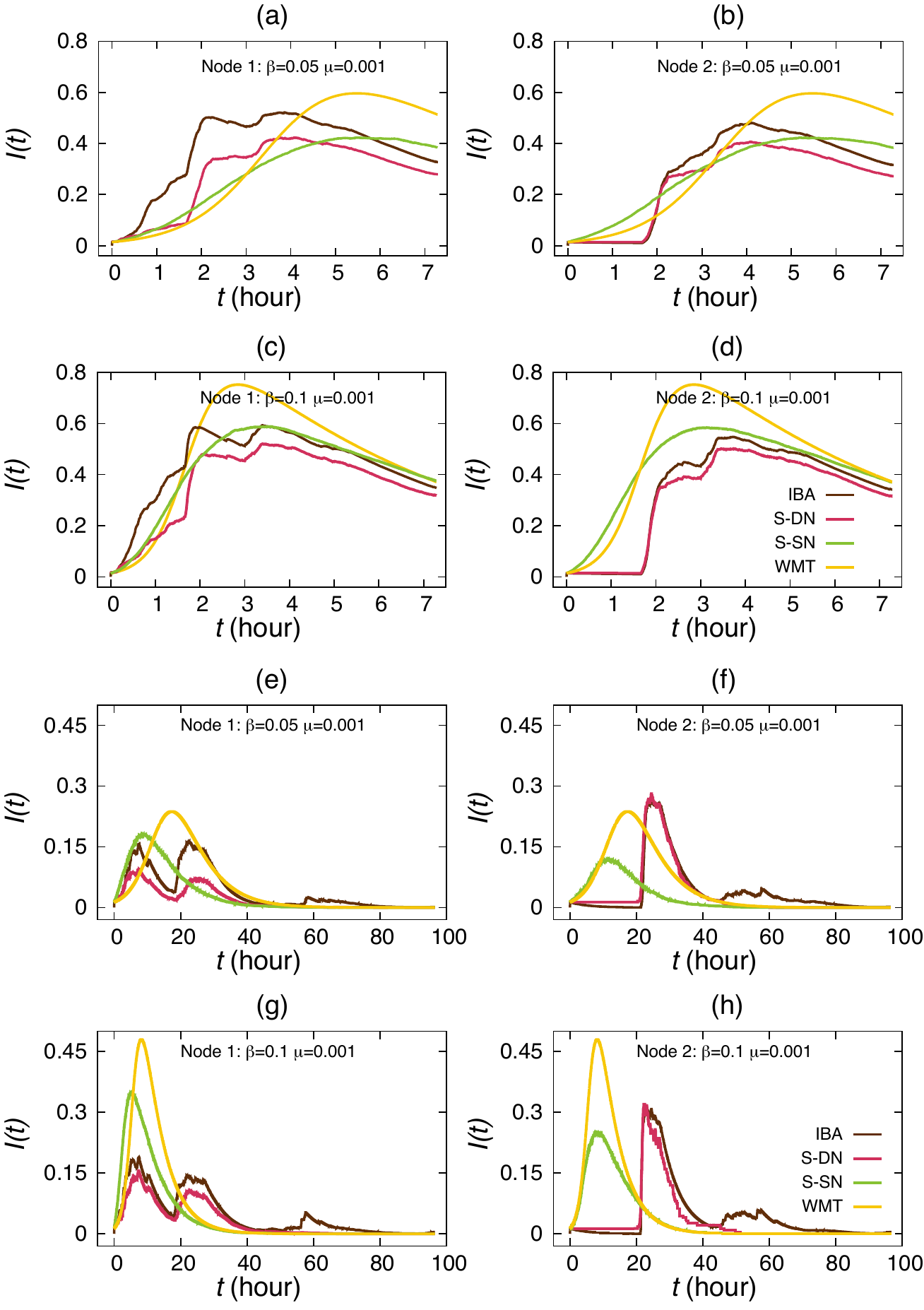}
\caption{\label{figS03}Accuracy of different approximation methods for the SPM and SPH data sets. The fraction of infected individuals as a function of time (i.e., the prevalence $I(t)$) for (a, b) SPM with $\beta= 0.05$ and $\mu=0.001$; (c, d) SPM with $\beta=0.1$ and $\mu=0.001$; (e, f) SPH with $\beta= 0.05$ and $\mu=0.001$; and (g, h) SPH with $\beta=0.1$ and $\mu=0.001$. The infection started at the seed individual (a, c, e, g) 1 or (b, d, f, h) 2.}
\end{figure}

\section*{Other datasets of dynamic contact networks}

We carried out some of the analyses presented in the main text to other data sets of dynamic contact networks to assess the generalisability of the results. These extra data sets correspond to face-to-face human interaction between visitors of a museum exhibition (SPM)~\cite{Isella11} and between patients and health-care workers in a hospital ward (SPH)~\cite{Vanhems2013} (Table~S\ref{tabS01}). As we show in the following, the results obtained for the conference dataset (SPC), used in the main text, qualitatively hold true for these data sets as well. 

\begin{figure}[h]
\centering
\includegraphics[scale=0.7]{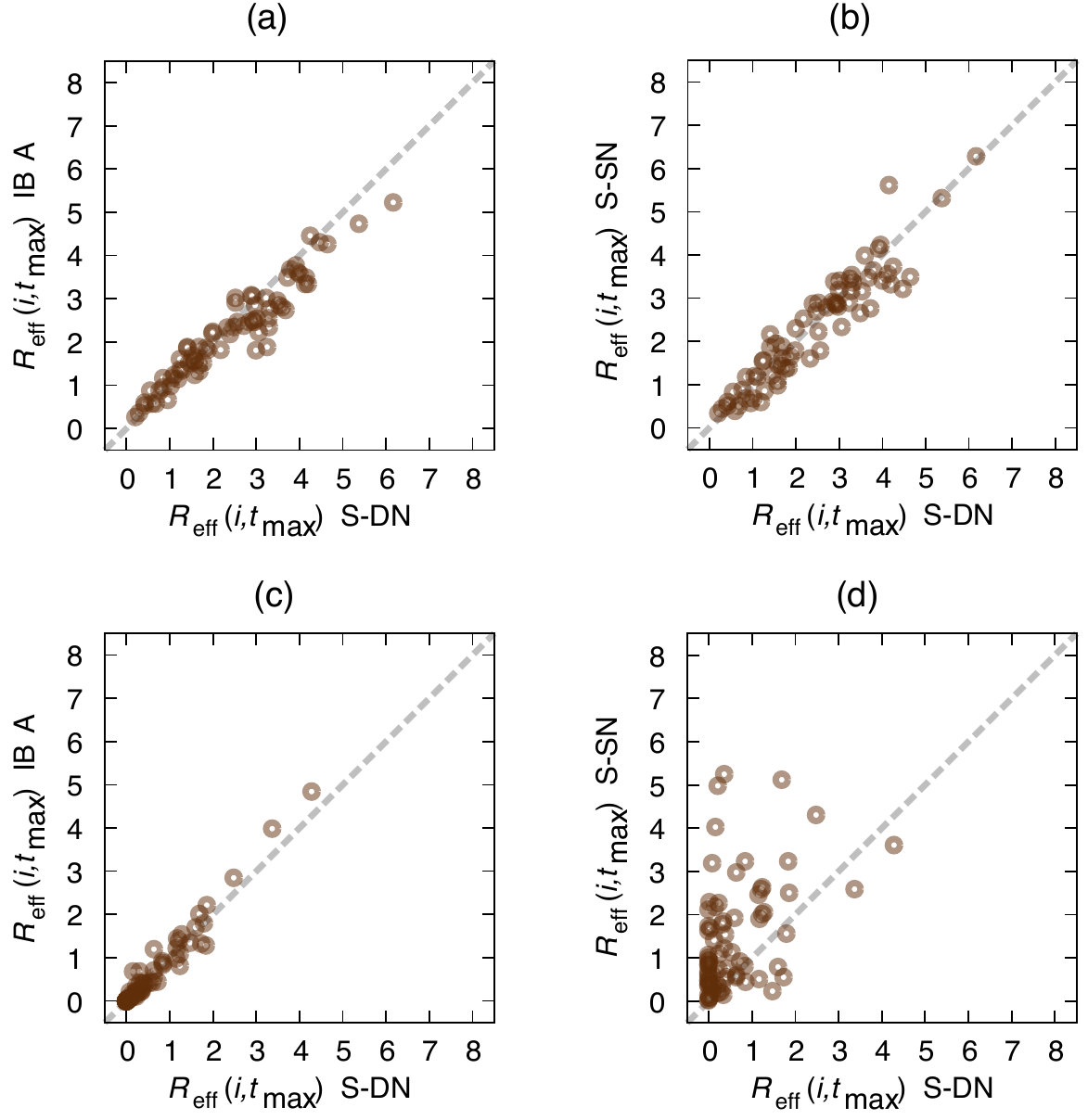}
\caption{\label{figS04}Individual effective reproduction number at time $t_{\text{max}}$. Comparison between (a) the S-DN and IBA for SPM; (b) the S-DN and S-SN for SPM; (c) the S-DN and IBA for SPH; and (d) the S-DN and S-SN for SPH. We set $\beta=0.05$ and $\mu=0.001$. Each circle represents an index individual $i$.}
\end{figure}

\begin{figure}[h]
\centering
\includegraphics[scale=1.3]{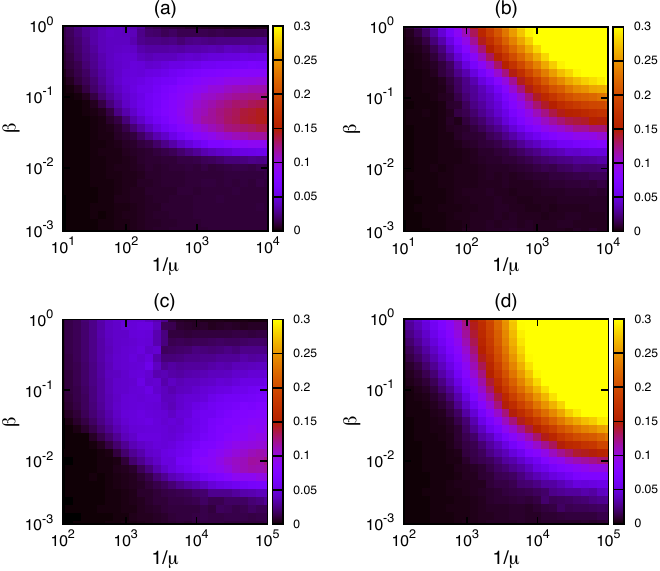}
\caption{\label{figS05}The root-mean-square RMS (equation~\eqref{eq:01}) between (a) the IBA and S-DN for the SPM network; (b) the S-SN and S-DN for the SPM network; (c) the IBA and S-DN for the SPH network; and (d) the S-SN and S-DN for the SPH network. Various combinations of infection probability $\beta$ and recovery probability $\mu$ are scanned. The colours represent the value of the RMS between the two analysed models.}
\end{figure}

\subsection*{Number of infected individuals}

Figure~S\ref{figS03} shows the evolution of the fraction of infected individuals, $I(t)$ (i.e., the prevalence), for SPM and SPH networks, given two different initial conditions and two combinations of epidemiological parameter values. We compare the estimation of the prevalence based on the IBA, S-DN, S-SN and WMT. In line with the results for SPC (main text), we observe a reasonable agreement between the IBA and S-DN. On the other hand, the S-SN and WMT are unable to reproduce the evolution of the prevalence observed in the S-DN. In the non-temporal models, the peak prevalence is also shifted either to earlier or later times depending on the combination of parameters. We also see that for both data sets, individual 2 (arbitrarily chosen) does not spread the infection before making the first contact according to the IBA and S-DN, a phenomenon not captured by non-temporal models such as the S-SN and WMT.

\begin{figure}
\centering
\includegraphics[scale=0.6]{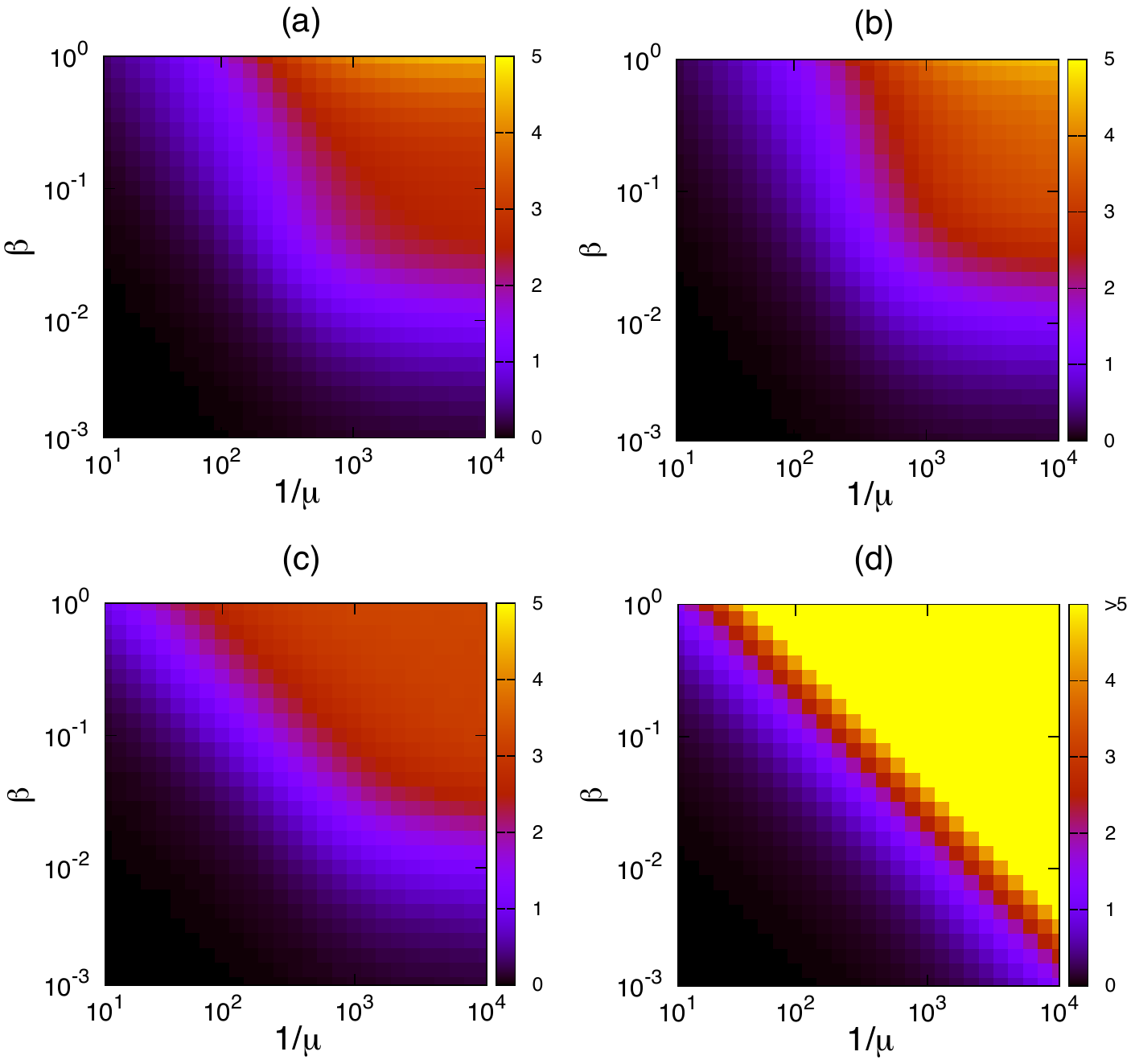}
\caption{\label{figS06}Effective reproduction number $R_{\text{eff}}(t_{\max})$ for various epidemiological parameters $\beta$ and $\mu$ for SPM data set. (a) IBA; (b) S-DN; (c) S-SN; and (d) WMT models. The colours represent the value of $R_{\text{eff}}(t_{\max})$.}
\end{figure}

\subsection*{Individual reproduction number}

For both data sets, the IBA is also accurate in approximating the S-DN in terms of the individual reproduction number (Fig.~S\ref{figS04}(a) and~S\ref{figS04}(c)). For the SPM network, we also observe an agreement between the S-DN and S-SN. This happens because the SPM data set contains only one day and thus the temporal heterogeneity created by the day-night cycle is not present in this data set. The SPH contact network yields a disagreement between S-DN and S-SN, in contrast to the results shown in the main text for SPC. In the SPH case, the S-SN tends to overestimate the individual reproduction number.

\begin{figure}[h]
\centering
\includegraphics[scale=0.6]{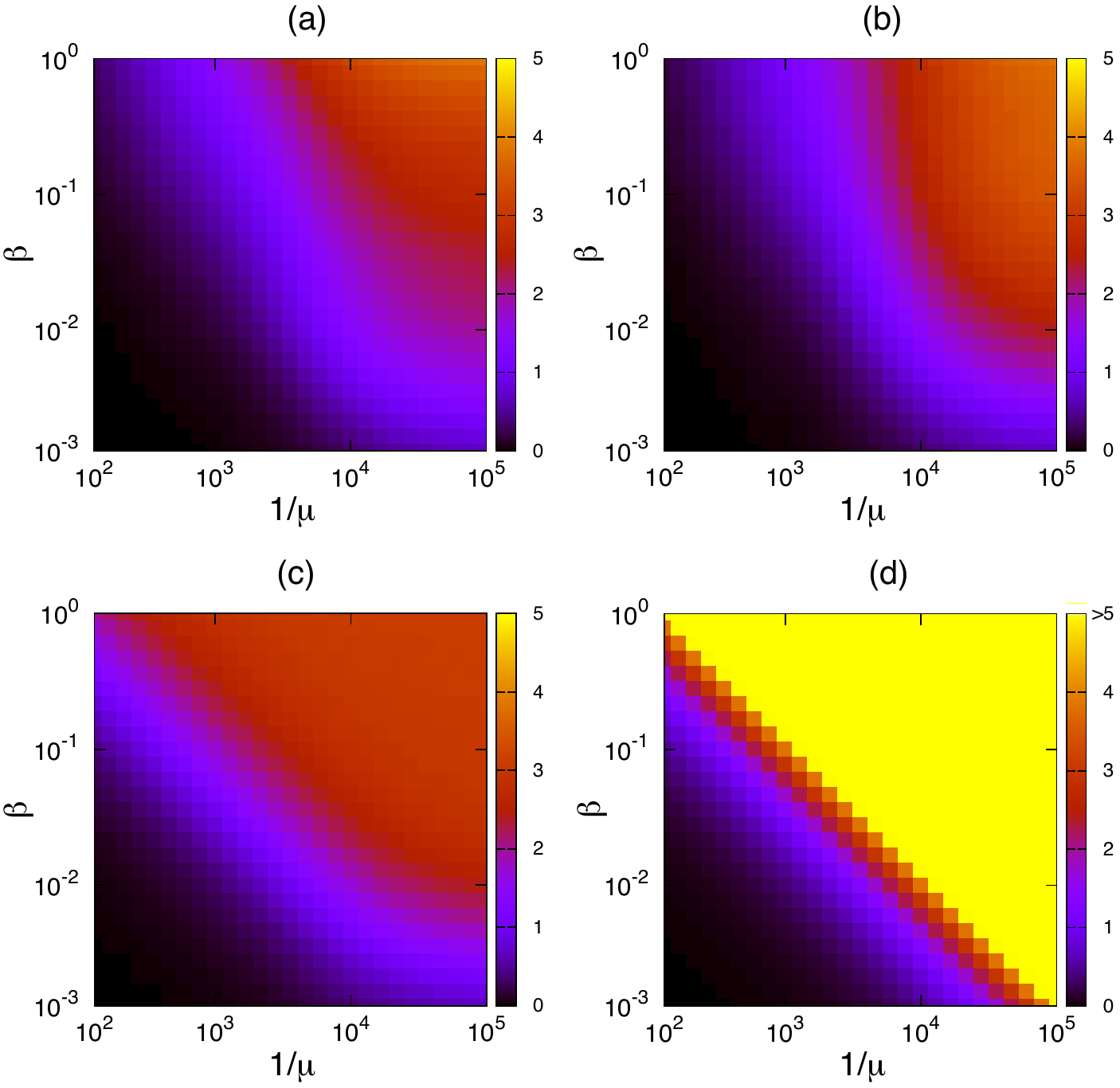}
\caption{\label{figS07}Effective reproduction number $R_{\text{eff}}(t_{\max})$ for various epidemiological parameters $\beta$ and $\mu$ for SPH data set. (a) IBA; (b) S-DN; (c) S-SN; and (d) WMT models. The colours represent the value of $R_{\text{eff}}(t_{\max})$.}
\end{figure}

\begin{figure}
\centering
\includegraphics[scale=0.55]{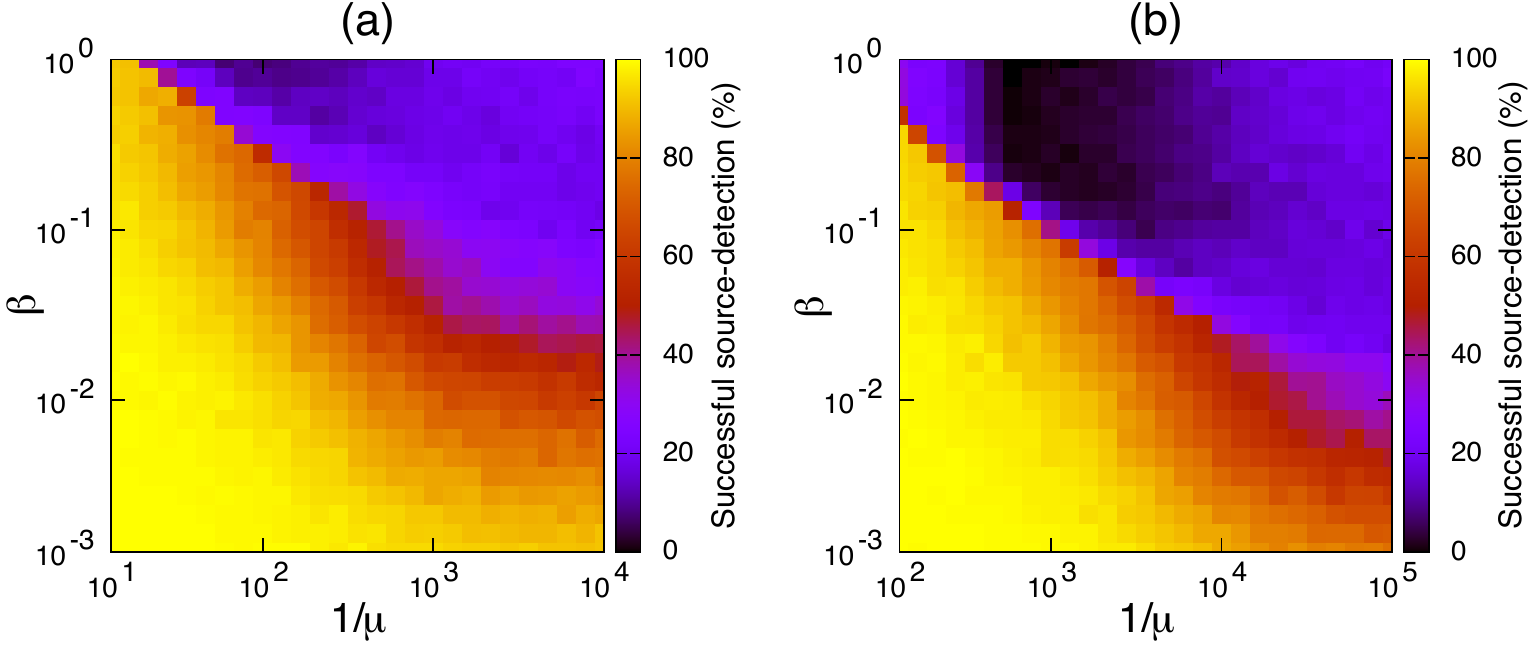}
\caption{\label{figS08} Percentage of successfully detected infection sources (estimated on the basis of 1000 randomly located sources) for various infection and recovery probabilities for (a) SPM and (b) SPH data sets. We set $t=t_{\text{max}}$. }
\end{figure}

\subsection*{Accuracy of the IBA}

We calculate the root-mean-square RMS deviation (equation~\eqref{eq:01}) for the SPM and SPH data sets (Fig.~S\ref{figS05}). We observe that the RMS between the IBA and S-DN is small for nearly all combinations of $\beta$ and $\mu$ for both data sets. On the other hand, the S-DN and S-SN show a relatively large disagreement in most of the $\beta$--$\mu$ parameter space, particularly for large $\beta$ and small $\mu$.

\subsection*{Effective reproduction number}

We calculate the effective reproduction number for the IBA, S-DN, S-SN and WMT models. The results agree between the IBA and S-DN in the full range of parameters studied (Fig.~S\ref{figS06}(a)-(b) for SPM and Fig.~S\ref{figS07}(a)-(b) for SPH). In contrast, the S-SN and S-DN do not agree for relatively large $\mu$ (Fig.~S\ref{figS06}(c) for SPM and Fig.~S\ref{figS07}(c) for SPH). We also observe no agreement between the WMT and S-DN in the entire range of parameters (Fig.~S\ref{figS06}(d) for SPM and Fig.~S\ref{figS07}(d) for SPH).

\subsection*{Source detection}

We perform the source-detection experiments using the SPM and SPH data sets. Similarly to results for the SPC data set, the IBA efficiently detects the source of infection given the states of individuals at time $t_{\max}$ and the past contact patterns (Fig.~S\ref{figS08}(a) for SPM and Fig.~S\ref{figS08}(b) for SPH).


\end{document}